\documentclass{pazhb}
\usepackage[hyperfootnotes=false]{hyperref}
\usepackage{color}
\usepackage{epsf}
\usepackage[utf8]{inputenc}
\usepackage[english, russian]{babel}
\usepackage[T2A]{fontenc}
\usepackage{indentfirst}
\usepackage{amsmath}
\usepackage{amsfonts}
\usepackage{amssymb}
\usepackage{epsf}
\usepackage{graphicx}
\usepackage{gensymb}
\usepackage{float}
\usepackage{longtable}
\usepackage{changes}
\usepackage{wrapfig}

\begin{document}

\newcommand{\obj}{BM~CrB}

\journalinfo{2023}{49}{3}{1}[13]
\UDK{///}

\title{Optical study of the polar BM~CrB in low accretion state}
\author{
A.I.~Kolbin\email{kolbinalexander@mail.ru}\address{1,2}, 
N.V.~Borisov\address{1}, 
A.N.~Burenkov\address{1}, 
O.I.~Spiridonova\address{1}, 
I.F.~Bikmaev\address{2,3}, 
M.V.~Suslikov\address{1,2}
\addresstext{1}{Special Astrophysical Observatory, Nizhnij Arkhyz, Karachai-Cherkessian Rep., 369167, Russia}
\addresstext{2}{Kazan (Volga-region) Federal University, 18 Kremlevskaya str., Kazan 420008, Russia}
\addresstext{3}{Academy of Sciences of Tatarstan Rep., 20 Baumana str., Kazan 420111, Russia}
}


\shortauthor{A.I. Kolbin, et al.}

\shorttitle{Optical study of BM~CrB}

\begin{abstract}
This paper presents a spectral and photometric study of the poorly studied polar {\obj}. Three states of the polar brightness and signs of transition from one-pole to two-pole accretion mode were found by an analysis of ZTF data. It is shown that the transition from the low state to the high state changes the longitude of the main accretion spot (by $\approx 17^{\circ}$) and increases its elongation (by $\approx 10^{\circ}$). The spectra contain Zeeman absorptions of the H$\alpha$ line which are formed at a magnetic field strength of $15.5\pm1$~MG. These absorptions are likely produced by a cold halo extending from the accretion spot at $\approx {^1/_4}$ of the white dwarf radius. Modeling of the behavior of the H$\alpha$ emission line shows that the main source of emission is the part of the accretion stream near the Lagrangian point L$_1$, which is periodically eclipsed by the donor star. The spectra exhibit a cyclotron component formed in the accretion spot. Its modeling by a simple accretion spot model gives constraints on the magnetic field strength $B=15-40$~MG and the temperature $T_e\gtrsim15$~keV.

\end{abstract}

\section*{Introduction}

Cataclysmic variables are close binary systems with orbital periods $P_{orb} \sim 1.4 - 9$~hours, which consist of a white dwarf (primary) and a cold main sequence star or brown dwarf (secondary) \citep{ Warner95, Hellier01}. The secondary fills its Roche lobe and loses a matter through the inner Lagrangian point L$_1$. In most cases, the material emitted by the donor forms an accretion disk around the white dwarf. However, when the accretor has a strong magnetic field ($B\sim 10-100$~MG), the ionized gas of the stream quickly reaches the stagnation region, where the dynamic pressure of the gas becomes equal to the magnetic pressure ($\rho v^2 = B^2/8 \pi$), and then flows along the magnetic lines without forming a disk. Systems of this type are called AM~Her type variables or polars. The gas trajectory in such systems can be divided into two components: ballistic, extending from the Lagrangian point L$_1$ to the stagnation region, and magnetic, where the ionized gas moves along the magnetic field lines in the direction of the white dwarf's magnetic poles. Unlike non-magnetic systems, polars have synchronization of the white dwarf's rotation and orbital motion ($P_{rot} = P_{orb}$, where $P_{rot}$ is the period of the white dwarf's rotation). By the interaction of the incident gas with the surface of the accretor the hot ($T\sim 10-50$~keV) accretion spots are formed. They are strong sources of X-ray bremsstrahlung, as well as polarized cyclotron radiation in the optical range. In long-term photometric observations of polars, high and low states are distinguished, differing in average brightness by several magnitudes. Switching between high and low states does not show any periodicity. Most likely, low states are formed due to the suppression of matter transfer by the local magnetic fields of the donor \citep{King98}. For a more detailed introduction to AM~Her type systems, we refer the reader to the review of \cite{Cropper90}.

The object {\obj} (SDSS J154104.67 +360252.9) was classified by \cite{Szkody05} as a polar in a study of cataclysmic variables detected by the Sloan Sky Survey. Their spectropolarimetry revealed features typical for polars: emission lines of hydrogen and neutral helium; strong line of ionized helium HeII~$\lambda$4686, comparable in intensity to H$\beta$; high degree of circular polarization (up to 9\%). Harmonics of the cyclotron line were also detected, whose position corresponded to a magnetic field strength of 33~MG. The same paper presents the {\obj} light curve, which is modulated with a period $P\approx 1.4$~h and has a bright ($\Delta V \approx 1.3$~mag) maximum lasting about half the period. It was assumed that the maximum is formed during the passage of the accretion spot across the disk of the white dwarf. 

The poor knowledge of {\obj} prompted us to perform a more detailed study of it using phase-resolved spectroscopy. In addition, unlike \cite{Szkody05}, our observations were carried out in a lower state {\obj}, which promised to obtain new information about the properties of the object. In addition to spectroscopic observations, we made photometric observations of {\obj} and also investigated the long-term behavior of the brightness based on data from the ZTF survey \citep{masci18}. 

\section{Observations and data reduction}
\label{section_obs}

\subsection{Spectroscopy} 

The set of {\obj} spectra was obtained with the 6m BTA telescope of the Special Astrophysical Observatory of the Russian Academy of Sciences using the SCORPIO focal reducer in the long-slit spectroscopy mode \citep{Afan}. The observations were made on the nights of April 25–26, 2022 and April 26–27, 2022 under good astroclimatic conditions (seeng $\approx 1.8-2.2''$). The VPHG550G volume phase holographic grating (550 strokes/mm) was used as a disperser, which at a slit width of 1.2$''$ provided spectral range coverage $\lambda = 3800-7300$~\AA\, with average resolution  $\Delta \lambda \approx 8$~\AA. On the first night 10 spectra were obtained with exposures of 600~sec, and on the second night --- 21 spectra with exposures of 300~sec. 

The spectral observations were processed using the IRAF software package\footnote{The IRAF astronomical data reduction and analysis software package is available at https://iraf-community.github.io.}. The bias images were subtracted from the spectral frames, and the correction for sensitivity microvariations of the device was performed based on the images of the flat-field lamp. Removal of traces of cosmic particles was performed using the LaCosmic algorithm based on the Laplacian analysis of images \citep{Dokkum}. The correction of geometric distortions and wavelength calibration of the spectra were performed using frames of the He-Ne-Ar lamp. The optimal extraction of the spectra \citep{Horne86} with subtraction of the sky background was performed. The spectrophotometric calibration was based on the observations of the AGK+81~266 standard. The fluxes {\obj} were corrected for the variable opacity of the atmosphere using the spectra of the neighboring star captured by the spectrograph slit. The barycentric Julian dates and barycentric corrections for the radial velocities where found for each spectrum.

\subsection{Photometry} 

Photometric observations of {\obj} were carried out on 1m Zeiss-1000 telescope of the Special Astrophysical Observatory of the Russian Academy of Sciences. The telescope was equipped with a nitrogen-cooled photometer with a $2$K$\times 2$K EEV 42-40 CCD. Additional photometric observations were made with the 1.5m RTT-150 telescope (Turkish National Observatory T\"UB\.ITAK) using the TFOSC instrument in photometer mode. In addition, several images of the {\obj} neighborhood were obtained before spectral observations on the BTA/SCORPIO to assess the state of the object. The log of photometric observations is presented in the table \ref{log_phot}.

The aperture photometry of {\obj} was performed using the tools of the IRAF package. The bias image was subtracted from the CCD frames, and correction for multiplicative errors was carried out using flat-field frames. The removal of traces of cosmic particles was carried out using the LaCosmic algorithm \citep{Dokkum}. The selection of the optimal aperture for photometry was carried out by minimizing the RMS brightness of control stars which were close in brightness to {\obj}.

\begin{table*}
\caption{The log of photometric observations of {\obj}. Telescopes and photometers involved in the observations, observation nights, duration of observations, number of obtained images ($N$), used filters, and exposure times ($\Delta t_{exp}$) are listed.}
\label{log_phot}
\begin{center}
\begin{tabular}{lccccc}
\hline
Telescope/         & Date,             & Duration,    & $N$	  & Filter & $\Delta t_{exp}$, 	\\
Photometer          & UT               & BJD-2459000          &     &        & sec \\ \hline
BTA/SCORPIO       & 25/26 Аpr 2022  & 695.470642--695.509242 & 8   & $V$       & 60 \\
Zeiss-1000/CCD    & 26/27 Apr 2022  & 696.435386--696.517802 & 43  & $V$       & 120 \\
BTA/SCORPIO       & 26/27 Apr 2022  & 696.443181--696.460126 & 6   & $V$       & 120 \\
RTT-150/TFOSC     & 07/08 May  2022  & 707.382490--707.483894 & 42  & $V$       & 180    \\
Zeiss-1000/CCD    & 28/29 May  2022  & 728.349864--728.420517 & 40  & $V$       & 30 \\
\hline
\end{tabular}
\end{center}
\end{table*}

\section{PHOTOMETRY ANALYSIS}

\subsection{ZTF photometry}

\begin{figure*}
  \centering
	\includegraphics[width=\textwidth]{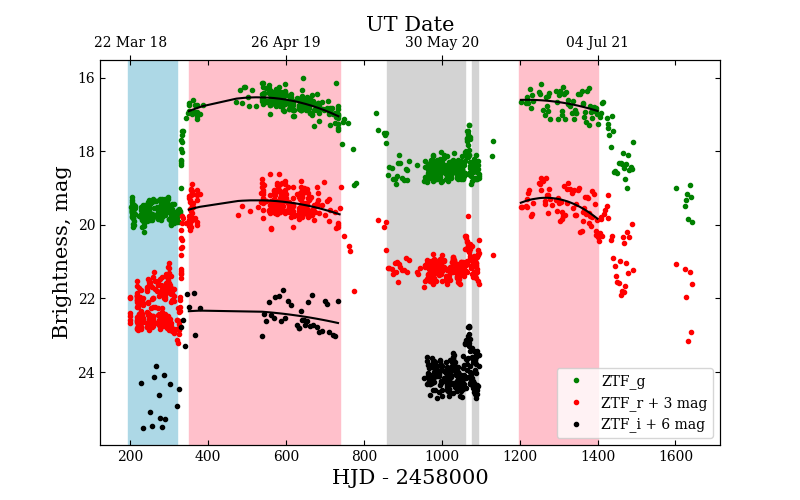}
\caption{The light curves of {\obj} from the ZTF survey in the $g$, $r$, $i$ bands. Areas shaded in pink, gray, and blue indicate high, intermediate, and low states, respectively. The data falling in these regions were used to analyze the light curves in three states. The black lines are second degree polynomials approximating the light curves in high state.}
\label{fig:phot_long}
\end{figure*}

The analysis of the long-term behavior of the brightness of {\obj} was carried by the data of ZTF survey \citep{masci18} covering approximately four years. The light curves of {\obj} in three photometric bands are shown in Fig. \ref{fig:phot_long}. It can be seen that during the observations the polar was in three states of brightness: low ($\langle g \rangle \approx 19.5$~mag), intermediate ($\langle g \rangle \approx 18.5$~mag), and high ($\langle g \rangle \approx 16.5$~mag).

\begin{figure*}
  \centering
	\includegraphics[width=\textwidth]{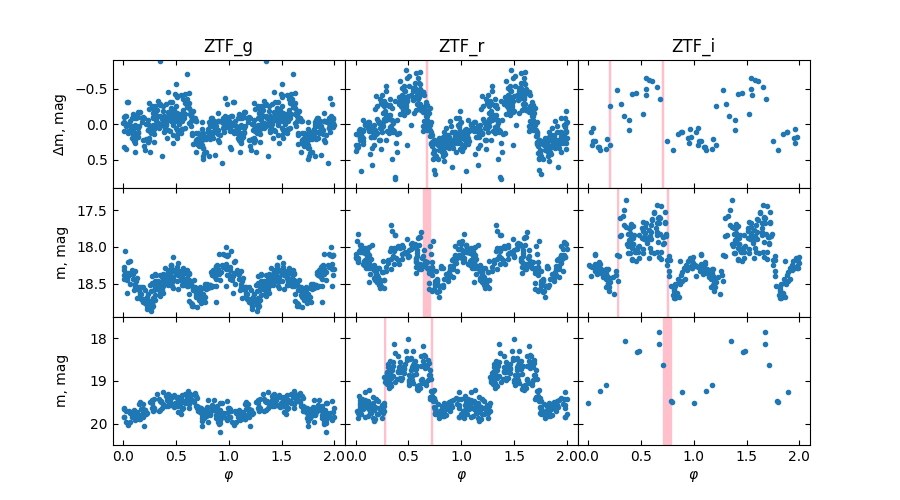}
\caption{Phase foldered light curves of {\obj} obtained from ZTF data. The polar states are listed from top to bottom: high, intermediate and low states. Photometric filters are listed from left to right ($g$, $r$, $i$). The data used to construct these light curves are shown in Fig. \ref{fig:phot_long}. The polar brightness in the high state is measured relative to the approximating polynomials shown in Fig. \ref{fig:phot_long}. The vertical stripes indicate the igress and egress of the bright phase; the widths of the stripes correspond to errors in determining the boundaries of the bright phase.}
\label{fig:lcs_phased}
\end{figure*}

The photometric period of {\obj} was refined by the Lomb--Scargle method, which was separately applied to different photometric bands and brightness states. The data used to analyze the behavior of the object in different states are marked in Fig. \ref{fig:phot_long} by shaded areas. Before examining the light curves in the high state, the trends approximated by parabolas (see Fig. \ref{fig:phot_long}) were subtracted. The average value of the period is $P = 84.0653 \pm 0.0002$~min. The light curves of {\obj} foldered with the found period in three states are shown in Fig. \ref{fig:lcs_phased}. These light curves were foldered according ephemeris
\begin{equation}
    HJD_{\min} = 2458862.097(1) + 0.0583786(1) \times E,
\label{ephem}
\end{equation}
where the zero epoch corresponds to the centre of the plateau in the $r$-band light curve in the low state. 

The light curves shown in Fig. \ref{fig:lcs_phased} alter their shape considerably when the state or the photometric band are changed. In the low state, in the $r$, $i$ bands, there is a dim plateau with a slight brightness variations ($-0.75 \lesssim \varphi \lesssim 0.25$), as well as a bright phase ($0.25 \lesssim \varphi \lesssim 0.75$), where object brightens by $\Delta r \approx \Delta i \approx 1.5$~mag. The bright phase occupies about half of the period and has a shape close to rectangular. In the $g$-band the bright phase also appears, but it is less pronounced (a brightening of about $0.5$~mag). It can be assumed that the bright phase occurs during the passage of the accretion spot accross the disk of the white dwarf. The bright phase in the $i$ band exhibits a two-hump structure. This effect is ordinary for polars and is often interpreted by cyclotron beaming (see e.g. \cite{Kolbin20}).

When {\obj} passes to an intermediate state, a second maximum appears near the phase $\varphi \approx 0$ (see Fig. \ref{fig:lcs_phased}).  It becomes dominant in $g$ filter. It is likely that the appearance of the second maximum is due to the second accretion spot formed near the second magnetic pole of the white dwarf. In the high state the highest brightness amplitude appears in $r$ band. The amplitude of the second maximum is significantly reduced. In the $i$ filter, the light curve has a pronounced primary maximum with an almost rectangular shape. The maximum is reduced when switching to short-wavelength filters. 

As shown in Fig. \ref{fig:lcs_phased}, there is a noticeable shift in the position of the bright phase ($0.25 \lesssim \varphi \lesssim 0.75$) with the change of average brightness. The boundaries of the bright phase were determined by approximating the light curve with a trigonometric polynomial. Those phases were taken as boundaries where the derivative of the trigonometric polynomial reached an extremum. The boundary position errors were found by the Monte-Carlo method. The shift of the right boundary of the bright phase in the $r$ filter between the high and low states is $\Delta \varphi = 0.047\pm0.006$, which corresponds to the shift of the spot longitude by $\Delta \psi = 17\pm 2^{\circ} $. A similar shift in the $i$ band is $\Delta \psi = 16\pm 7^{\circ}$. A more accurate value of the phase shift in the $i$ band can be obtained between the intermediate and high states. For the right spot boundary, it is $\Delta \psi = 17 \pm 2^{\circ}$, and for the left one $\Delta \psi = 27\pm2^{\circ}$, i.e. during the transition from intermediate to high the spot state is stretched in longitude by $\approx 10^{\circ}$. A similar behavior of the light curve was observed by \cite{Schwope15} for the polar V808~Aur. The shift of the accretion spot is naturally explained by the increase in the average brightness of the polar with an increase in the accretion rate. With an increased accretion rate, the accretion stream should have a higher density and be later captured by magnetic field lines (the Alfven radius becomes smaller). In this case, the spot is formed at higher longitudes (if the longitudes are counted in the direction of the donor's orbital motion), and the bright phase sets in earlier, which is observed in the {\obj} light curves.

\subsection{RTT-150 and Zeiss-1000 photometry}
\label{section:phot_our}

\begin{figure}
  \centering
	\includegraphics[width=\columnwidth]{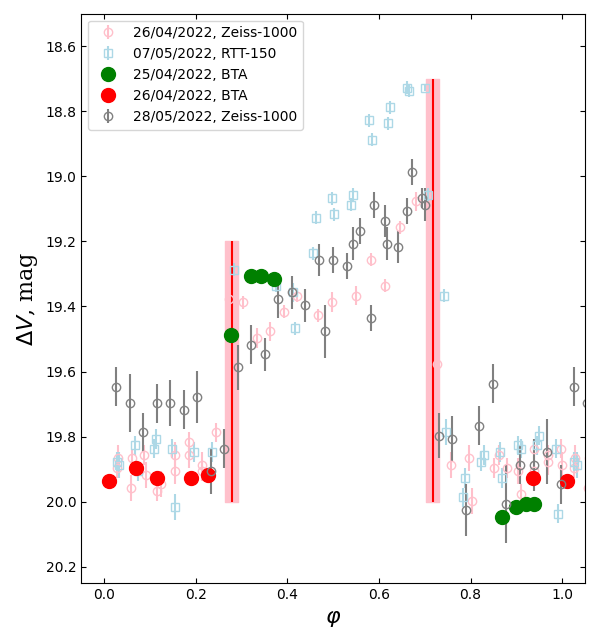}
\caption{The light curves of {\obj} obtained with the Zeiss-1000, RTT-150 and BTA telescopes. The vertical lines indicate the moments of egress and ingress of the bright phase.}
\label{fig:lcs_our}
\end{figure}

The {\obj} light curves obtained with the RTT-150 and Zeiss-1000 in the $V$ band are shown in Fig. \ref{fig:lcs_our}. They show a plateau with a constant brightness level and a bright phase that extends for about half the period and is formed by the passage of the accretion spot on the disk of a white dwarf. Brightness on the plateau $V\approx 19.8$~mag indicates a low state of {\obj} at the time of observations. Ingress and egress of the bright phase are abrupt. After a sharp ingress the brightness gradually increases by $\Delta V \approx 0.4$~mag.

Based on the Zeiss-1000 observations on May 28, 2022, which have the highest time resolution, the duration of the bright phase is $\Delta \varphi_{BP} = 0.44 \pm 0.02~P_{rot}$, where $P_{rot}$ is the rotational period if the white dwarf ($P_{rot} = P_{orb}$ in polars). Since $\Delta \varphi_{BP} < {^1/_2} P_{rot}$, the accretion spot must be located on the white dwarf's hemisphere, turned away from the observer. 

The light curves of the RTT-150 and Zeiss-1000 were also overlaid with the brightness measurements obtained with the BTA/SCORPIO just before the spectral observations. Obviously, these measurements are in good agreement with the RTT-150 and Zeiss-1000 data and allow us to state that the polar was in a low state during spectral observations as well. 

\section{Zeeman splitting}
\label{section_spec}

\begin{figure*}
  \centering
	\includegraphics[width=\textwidth]{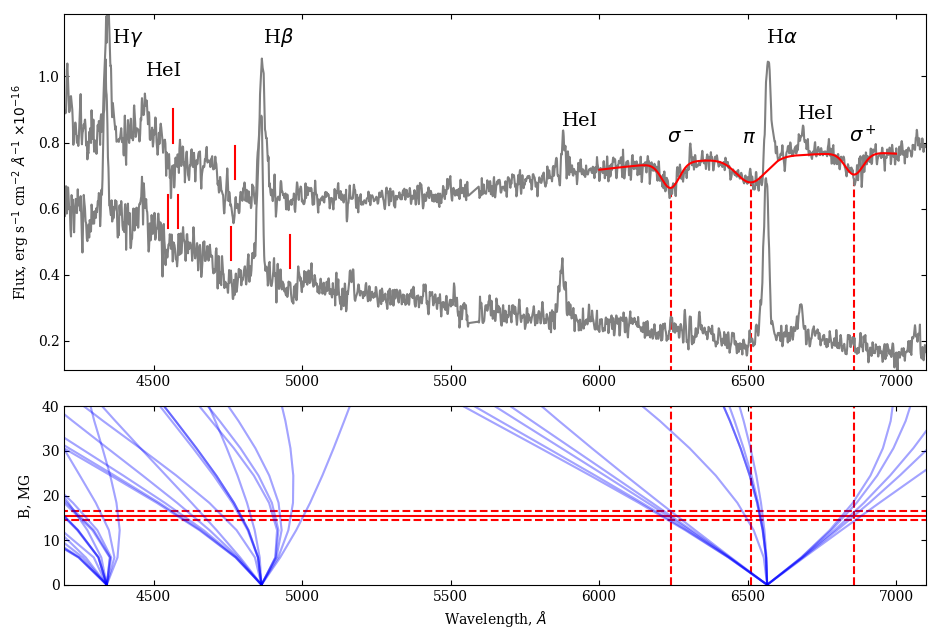}
\caption{The upper panel shows the averaged {\obj} spectra obtained in the bright phase and the plateau phase. The red line shows the approximation of the spectral region with Zeeman components of the H$\alpha$ line by a combination of a polynomial and gaussians. The vertical dotted lines indicate the position of the Zeeman components of H$\alpha$ line. The vertical dashes indicate the splitting components of the H$\beta$ line. The bottom panel shows a diagram of the positions of the Zeeman splitting components of the H$\alpha$, H$\beta$ and H$\gamma$ lines for a magnetic field strength range of $0-40$~MG. The found value of the magnetic field $B=15.5\pm1$~MG is indicated by the horizontal lines in this diagram.}
\label{fig:specs_aver}
\end{figure*}

The {\obj} spectra can be divided into two groups. The first group corresponds to the bright phase in the light curve ($0.25 \lesssim \varphi \lesssim 0.75$), and the second group corresponds to the low brightness plateau phase ($-0.75 \lesssim \varphi \lesssim 0.25$). The averaged spectra of the bright phase and the plateau are shown in Figs. \ref{fig:specs_aver}. The spectra of the first group show a wide hump in the red part of the spectrum, while the second group of spectra has a monotonous increase in the continuum towards the short wavelength region. The hump in the red part is due to the cyclotron radiation of the accretion spot passing through the visible disk of the white dwarf and giving a maximum brightness in the long-wavelength region. The observed behavior of the spectra is in good agreement with the low-state multipassband ZTF photometry (see the previous section), where there is an increase in the brightness amplitude when going from the $g$ filter to the $r$ and $i$ filters. 

Absorptions at wavelengths of $6240$, $6510$ and $6858$~\AA\, are observed in the bright phase. Their position was determined by approximating a section of the spectrum by a combination of an algebraic polynomial and three gaussians. Note that to measure the wavelength of a feature near $6858$~\AA\, the {\obj} spectrum was divided by the spectrum of the standard star to remove the telluric Fraunhofer line B. These absorptions can be interpreted as the $\sigma^-$, $\pi$, and $\sigma^+$ components of the Zeeman splitting of the H$\alpha$ line. It is notable that they are present only in the bright phase and disappear in the plateau phase. This suggests that the absorptions are formed in the cold halo around the accretion spot. Similar behavior of the Zeeman components of H$\alpha$ is demonstrated by the polars V834~Cen \citep{wickra87}, EP~Dra \citep{schwope97b}, BS~Tri \citep{Kolbin22}. 

To determine the magnetic field strength, we used the code of \cite{Schimeczek14}, which calculates the atomic states of the hydrogen in strong magnetic fields. We calculated the energies of the hydrogen states and the wavelengths of the allowed transitions for the range of magnetic field strengths of $0-40$~MG. The resulting splitting diagram for the H$\alpha$, H$\beta$ and H$\gamma$ lines is shown in the bottom panel of Fig. \ref{fig:specs_aver}. The position of the Zeeman components of H$\alpha$ is in the best agreement with the magnetic field strength $B_{halo}=15.5\pm1$~MG. Note that the magnetic field shows no variability within $\Delta B = 0.5$~MG during the bright phase. \cite{Szkody05}  estimated the magnetic field strength in the accretion spot at  $B_{cyc} \approx 33$~MG by measuring the wavelengths of cyclotron harmonics. Then, assuming a dipole magnetic field, the size of the halo is $(B_{halo}/B_{cyc})^{-1/3} -1 \approx 0.28$~of the white dwarf radius.

Besides the H$\alpha$ line splitting, the {\obj} spectra show the Zeeman splitting of the H$\beta$ line. It is noticeable both in the bright phase and in the plateau phase. In the plateau phase, the $H\beta$ absorption components are probably formed in the photosphere of the white dwarf, while in the bright phase the halo lines may overlap. Unfortunately, from the position of the H$\beta$ components it is impossible to make unambiguous estimates of the magnetic field strength. This problem could be solved by modelling the spectra of magnetic white dwarfs.

\section{Emission lines}

The set of emission lines observed in the {\obj} spectra is typical for cataclysmic variables. The most strong lines are the hydrogen lines of the Balmer series H$\alpha$, H$\beta$, H$\gamma$ (see Fig. \ref{fig:specs_aver}). There are also lines of the neutral helium HeI $\lambda$4471, $\lambda$5876, $\lambda$6678, $\lambda$7065. The interesting is the absence of HeII~$\lambda$4686 line, which is often found among variables of the AM~Her type and is often used as an indicator of the magnetization of a white dwarf in the cataclysmic variable \citep {Silber92}. This property can be interpreted by accretion state, when the accretion spot emits few hard photons ($\lambda \le 228$~\AA) and the accretion stream is too rarefied to form a strong HeII~$\lambda$4686 line. Note that in \cite{Szkody05} the intensity of the HeII~$\lambda$4686 line is comparable to the H$\beta$ line, and the {\obj} state was higher (the brigthness at the plateau is $V\approx 18.7$~mag) compared to with the present work (the brigthness at the plateau is $V\approx 19.8$~mag).

\begin{figure*}
  \centering
	\includegraphics[width=\textwidth]{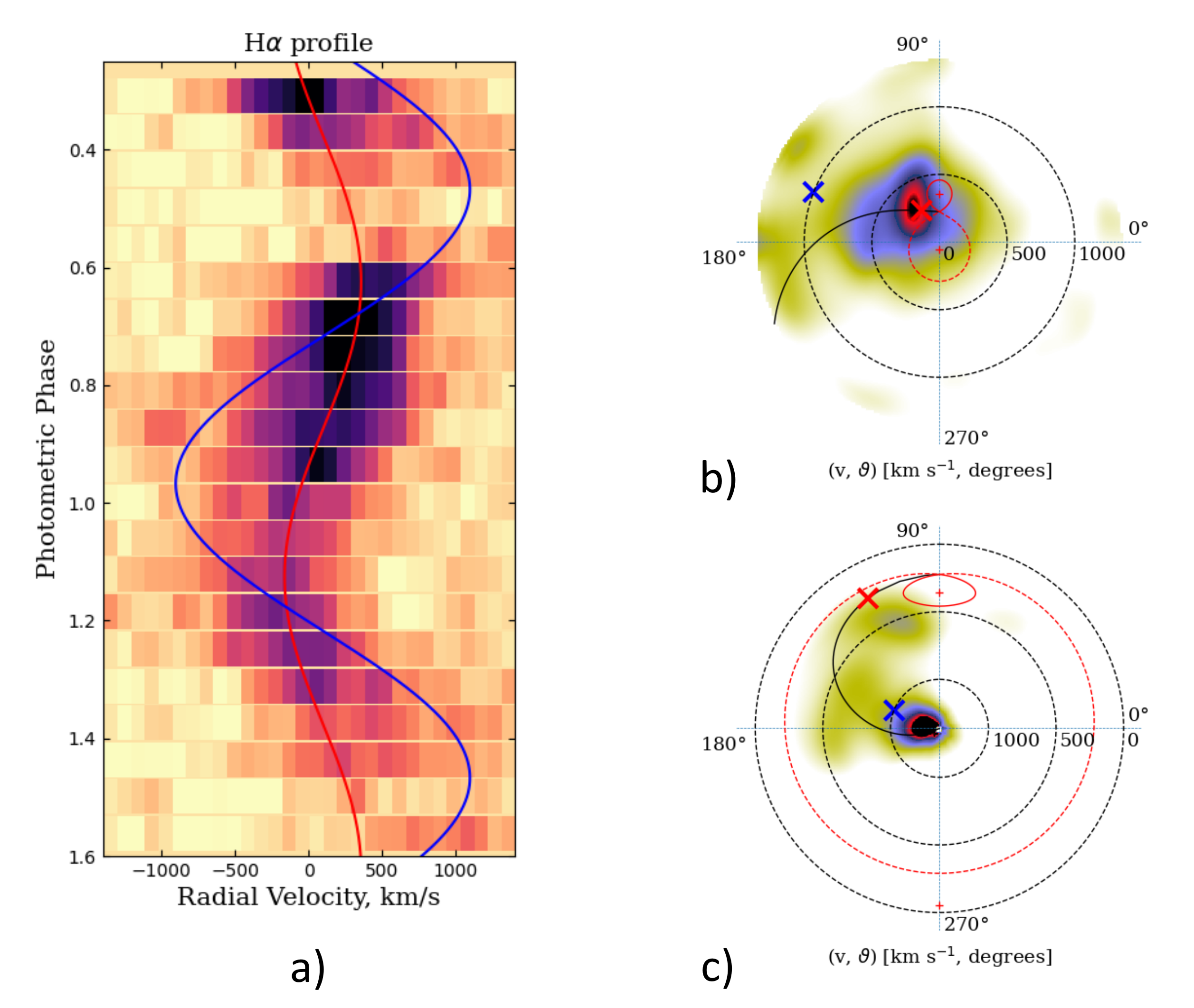}
\caption{
Dynamic spectrum of the H$\alpha$ line (a) and corresponding Doppler tomograms in standard (b) and inside-out (c) projections. The dynamic spectrum is superimposed with the radial velocities of the H$\alpha$ components derived by the decomposition of the profiles by Gaussians. The red line shows the radial velocity curve of the main component, the blue one shows the base component. The tomograms are superimposed with the Roche lobe velocities of the primary (closed red dotted line) and secondary components (closed red continuous line), particle velocities on the ballistic trajectory (solid black line). The position of the sources responsible for the formation of the main component of the line profile (bold red cross) and the broad base component (bold blue cross) is indicated.}
\label{fig:dyn_tom}
\end{figure*}

The dynamic spectrum of the H$\alpha$ line is shown in Fig. \ref{fig:dyn_tom}. Other emission lines behave similarly, but their dynamics is not presented in this paper due to high noise levels. The radial velocity variability modulated with the orbital period is evident. The line profiles were approximated by a set of Gaussians whose radial velocities varied as $\sim K \sin(2\pi(\varphi-\varphi_0))$, where $\varphi_0$ is the zero phase. Gaussian widths were independent of phase, and amplitudes were fitted to each profile. It turned out that the overwhelming majority of H$\alpha$ radiation is described by a single Gaussian with a radial velocity amplitude $K \approx 260$~km/s. For a better description of the dynamic spectrum, a weaker base component with an amplitude of $K \approx 1000$~km/s was introduced. The radial velocities of the line components are shown in Fig. \ref{fig:dyn_tom}. Note that the behavior of emission lines in polars is quite complex and is often described by a larger number of components (see \cite{Schwope97}). However, due to the high noisiness of our spectra and the low spectral resolution, the proposed model satisfactorily describes the observations. Although after subtracting the model profiles from the observed ones, a weak residual radiation is observed, the addition of one more Gaussian to describe the dynamic spectrum makes the solution ambiguous and strongly dependent on the initial guess.

In Fig. \ref{fig:dyn_tom} Doppler tomograms reconstructed from the H$\alpha$ dynamic spectrum are also shown (for details on Doppler tomograms see \cite{Marsh16}, \cite{Kotze15}, \cite{Kotze16}). Doppler tomograms are maps of the distribution of emission regions in velocity space projected onto the orbital plane. Each point of this space is defined by polar coordinates $v$ and $\vartheta$. The coordinate $v$ is the modulus of velocity measured from the center of mass of the binary system, and $\vartheta$ is the angle between the velocity vector of the radiating particle and the line connecting the centers of mass of the stellar components. Doppler tomography was performed using the program code of \cite{Kotze15}, which implements the maximum entropy method. The tomograms are presented both in the standard projection, where the modulus of velocity $v$ increases from the tomogram center to the periphery, and in the inside-out projection, where $v$ increases in the opposite direction. Since we do not have orbital ephemeris for {\obj}, the tomograms are rotated by an arbitrary angle $\vartheta$. It was assumed that the zero orbital phase (corresponding to the maximum proximity of the donor to the observer) happens to the photometric phase $\varphi = 0.5$, according to the BM~CrB model described below. The distribution of emitting regions in the presented tomograms is typical for systems of the AM~Her type: there is a low-velocity emission region, which passes into a high-velocity region with increasing polar angle. The tomogram shows the regions of formation of the main component and the base component, which were isolated from the dynamic spectrum by fitting the Gaussians. The deviation of the position of the main component from the local maximum on the tomogram in the inside-out projection is due to an increase in the blurring of details when moving from the center to the periphery of the tomogram (see \cite{Kotze15} for details). The same effect appears for the base component in the standard projection. 

An interesting feature in the behavior of emission lines is their disappearance in the phase interval $0.4 \lesssim \varphi \lesssim 0.6$. Apparently, this effect is associated with the eclipse of the emission source by the donor. The width of the eclipse is $\Delta \varphi \approx 0.17$, which indicates the formation of emissions at some distance from the donor rather than on its surface (for example, due to irradiation effect). If the emission is formed on the surface of a star then the minimum duration of the eclipse is achieved when the emission region is compact, located at the Lagrangian point L$_1$, and the inclination of the orbital plane is $i=90^{\circ}$. However, even in this case, a very small ans uncharacteristic for cataclysmic variables mass ratio $q=M_2/M_1\approx 0.0016$ would be required to obtain the observed width of the eclipse. The behavior of the radial velocity also calls attention to itself. If the emission source is an irradiation region, symmetrical with respect to the axis connecting the centers of mass of the components, then the center of the eclipse would correspond to a zero radial velocity of the orbital motion. However, in reality, it is ahead of the central phase of the eclipse by $\Delta \varphi \approx 0.08$.

\begin{figure}
  \centering
	\includegraphics[width=\columnwidth]{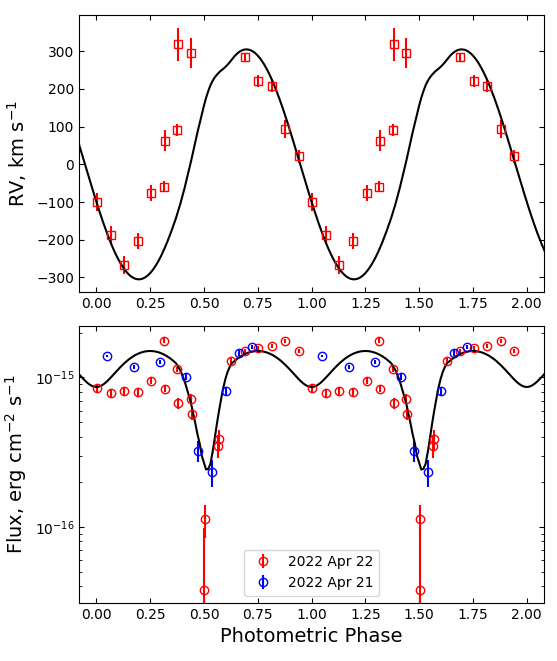}
\caption{Radial velocity curve (upper panel) and flux curve (lower panel) of the main component of the H$\alpha$ line. Solid lines show the corresponding theoretical curves.}
\label{fig:rvs}
\end{figure}

The radial velocity curve and the flux curve of the main emission source of the H$\alpha$ line are shown in Fig. \ref{fig:rvs}. These curves were obtained after subtracting from the spectra the weak base component, which was previously separated by the decomposition of the dynamic spectrum by Gaussians. The behavior of the H$\alpha$ line in {\obj} is very similar to the behavior of the emission lines in the polars HU~Aqr \cite{Schwope97} and QQ~Vul \cite{Schwope00}. A narrow eclipse is also observed in the flux curves of the HU~Aqr and QQ~Vul emission lines. Moreover, the flux curve in {\obj} shows little evidence of double-hump, with the left-hand hump being more intense than the right-hand one. A similar phenomenon is present in HU~Aqr and QQ~Vul. Based on the analysis of the high-resolution spectra in HU~Aqr and QQ~Vul, it was shown that most of the flux in the emission lines is formed in the accretion stream.

\begin{figure}
  \centering
	\includegraphics[width=\columnwidth]{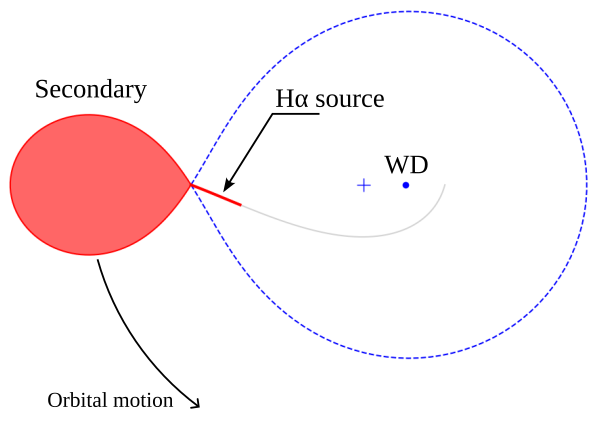}
\caption{The model of {\obj} obtained from the modeling of radial velocities and fluxes of the main component of the H$\alpha$ line. The figure shows: a donor star filling its Roche lobe, a white dwarf (WD), the Roche lobe of a white dwarf (dashed line), the ballistic trajectory of a particle emitted from the Lagrangian point L$_1$ (gray line), the source of the main component of the H$\alpha$ line.}
\label{fig:model}
\end{figure}

To test the hypothesis of the formation of {\obj} emissions in an accretion stream, we have developed a simple model of a semi-separated binary system. The model includes a secondary component (that fills its Roche lobe) and a part of the accretion stream flowing out of the Lagrangian point L$_1$. It was assumed that the source of the main component of H$\alpha$ is located near the donor before the stagnation region, where the stream trajectory can be considered to be ballistic. This assumption also saves us from ambiguities in the solution that could arise by introducing unknown parameters that determine the magnetic trajectory of the stream. The stream trajectory was calculated by solving the restricted three-body problem \citep{Flannery75}. The cross section of the accretion stream was assumed to be negligible. The stream was divided into small elements, which were considered to be emission sources in the H$\alpha$ line. The flux curve and the radial velocity curve were calculated by integrating the local intensities and radial velocities over the stream, respectively. The integration took into account the eclipse of the stream elements by the donor and the projection onto the plane of the sky (i.e., the radiation intensity was assumed to be proportional to $\sin \gamma$, where $\gamma$ is the angle between the line of sight and the stream element). The latter means that the stream was assumed to be optically thick. The best fit was obtained for orbital inclination $i=55^{\circ}$, white dwarf mass $M_1 = 0.65$~M$_{\odot}$, and elongation of the bright part of the accretion stream $\alpha = 7^{\circ} $ counted from the center of the accretor from the direction to the Lagrange point L$_1$. The derived model of the system projected onto the orbital plane is shown in Fig. \ref{fig:model}. The comparison of the observed radial velocity curve and H$\alpha$ flux curve with the corresponding model curves is shown in Fig. \ref{fig:rvs}. One can see a satisfactory description of the flux curve, as well as a section of the radial velocity curve in the phase range $0.7 \lesssim \varphi \lesssim 1.2$. In the interval $0.2 \lesssim \varphi \lesssim 0.5$, the theoretical radial velocities are significantly (by $\Delta v \approx 100$~km/s) lower than the observed ones. Most likely, this discrepancy is due to the poor separation of the line profile into components, which is inevitable for low resolution spectra. The analyzed profiles could contain the emission of stream regions closer to the white dwarf, which are better manifested during the period of reduced brightness of the main emission source. The particle velocities of the ballistic trajectory calculated for the noted parameters of the system were superimposed on the Doppler tomogram in Fig. \ref{fig:dyn_tom}. It can be seen that the H$\alpha$ emission sources are distributed near the trace of the ballistic trajectory.

\section{Cyclotron spectra}

The simple model of an accretion spot is often used to study the cyclotron spectra of polars (see e.g. \cite{Campbell08, Kolbin19, Beuermann20}). This model assumes temperature and density homogeneity of the spot and is determined by four parameters: magnetic field strength $B$; electron temperature $T_e$; the angle between magnetic line and line of sight $\theta$; plasma parameter $\Lambda = \omega^2_p \ell / \omega_c c$, where $\omega_c = eB/m_e c$ is the cyclotron frequency, $\omega_p$ is the plasma frequency, $\ell$ is the geometric depth of the spot along the line of sight. The high Faraday rotation presented in polar's accretion spots reduces the solution of the polarized radiation transport equation to two independent equations for ordinary and extraordinary waves. Outgoing radiation intensities for the ordinary ($o$) and extraordinary ($e$) polarization modes are found as
\begin{equation}
    I_{o,e} = I_{RJ} (1 - \exp(-\alpha_{o,e} \Lambda)),
\end{equation}
where $I_{RJ} = k_B T_e \omega^2 / 8 \pi^3 c^2$ is Rayleigh-Jeans intensity per polarization mode ($k_B$ is Boltzmann's constant), $\alpha_{o,e}$ are cyclotron absorption coefficients in $\omega^2_p/\omega_c c$ units. The coefficients $\alpha_{o,e}$ in this work were calculated according to the method of \cite{Chan81}. The total radiation intensity is $I = I_o + I_e$.

\begin{figure}
  \centering
	\includegraphics[width=\columnwidth]{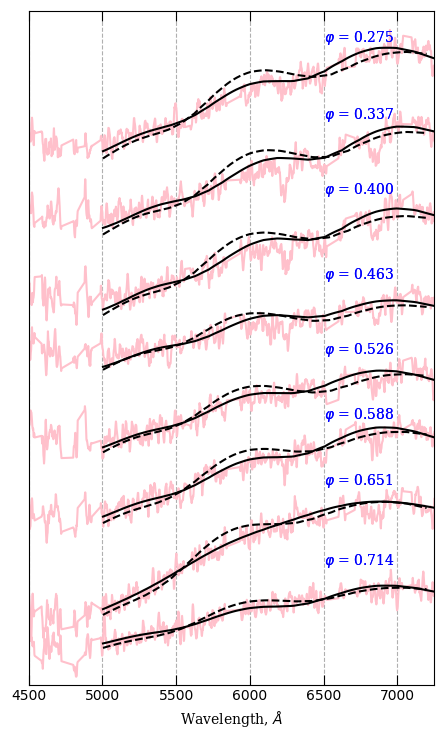}
\caption{Cyclotron spectra of {\obj} (pink lines) and fitted model spectra for temperatures $T_e=10$~keV (dashed black lines) and $T_e=20$~keV (solid black lines). The phases of the exposure midpoints of the observed spectra, found from the ephemeris (\ref{ephem}), are indicated.
}
\label{fig:cyc_spec}
\end{figure}

As noted in the section ``Photometry Analysis'', the accretion spot passes over the disk of a white dwarf during the bright phase ($0.25 \lesssim \varphi \lesssim 0.75$). Outside this phase, the light curves show a plateau, where, apparently, the radiation of the white dwarf dominates. To extract the spectrum of the accretion spot, we approximated the average plateau spectrum with a low-degree polynomial and subtracted it from the spectra of the bright phase. The spectra of the accretion spot obtained in this way are shown in Fig. \ref{fig:cyc_spec}. The flux in all spectra increases with increasing wavelength, and the presence of harmonics of the cyclotron line is not obvious. There is a sligt variability in the shape of the spectra, which is usually associated with changes in the visibility conditions of the accretion spot, i.e. with a change in the angle $\theta$ and the plasma parameter $\Lambda$.

The cyclotron spectra were approximated by varying the magnetic field strength $B$, the temperature $T_e$, the direction of the magnetic lines given by the angle $\theta$, and the plasma parameter $\Lambda$. The magnetic field strength and temperature were considered independent of the phase $\varphi$ and were found by fitting the entire set of cyclotron spectra. The angle $\theta$ and the plasma parameter $\Lambda$ were adjusted to each spectrum individually and varied in the ranges $\theta \in [0^{\circ}, 90^{\circ}]$, $\Lambda \in [2 ,8]$ (the last range covers the observed values of $\Lambda$ in polars and their theoretical limits, see, for example, \cite{Woelk96}). The wavelength range used for modelling was $\lambda = 5000-7100$~\AA, where the contribution of cyclotron radiation is largest and there is no addition from the second order of the diffraction grating.

A satisfactory description of the spectra is achieved for temperatures $T_e \gtrsim 15$~keV. The optimal values of the magnetic field strength are about $B=30$~MG within the temperature range $T_e=15-30$~keV. However, there is agreement with observations within the observational errors for the range $B=15-40$~MG. The resulting range of magnetic field strength does not contradict the estimate $B=33$~MG obtained by \cite{Szkody05}. However, the absence of pronounced cyclotron harmonics allowed us to estimate the parameters only with a high uncertainty. Note that a smooth change in the angle $\theta$ and the parameter $\Lambda$ was not revealed. The Fig. \ref{fig:cyc_spec} shows a comparison of the observed spectra with the theoretical spectra, which correspond to the optimal (for the least squares method) solutions for temperatures of 10 and 20~keV.



\section{Сonclusion}

We have performed an optical study of the poorly studied polar {\obj}, which was in a low brightness state during our observations. Below we present our main results.

An analysis of the long-term photometry obtained by the ZTF survey demonstrates three states of {\obj} brightness, which we called low ($\langle g \rangle \approx 19.5$~mag), intermediate ($\langle g \rangle \approx 18.5$~mag) and high ($\langle g \rangle \approx 16.5$~mag). The simplest light curve is observed in the low state, where one can distinguish a phase with a reduced, almost constant brightness (plateau phase), as well as a bright phase corresponding to the emergence of an accretion spot onto the disk of a white dwarf. The duration of the bright phase indicates accretion near the pole of rotation, directed away from the observer. With an increase in the state of brightness, signs of accretion onto the second magnetic pole appear. A change in the longitude of the main accretion spot ($\Delta \psi \approx 17^{\circ}$) and its extension (by about $10^{\circ}$ in longitude) during the transition from a low state to a high one are traced.

The spectra of the bright phase show absorption Zeeman components of the H$\alpha$ line, which are formed in a magnetic field with a strength of $B = 15.5 \pm 1$~MG. Since these absorptions are visible only in the bright phase, it can be assumed that they are formed in a cold halo around the accretion spot. If we accept the estimate of the magnetic field strength in the accretion spot $B=33$~MG \citep{Szkody05}, then the size of the halo should be $\approx 0.28$ of the white dwarf radius. In the plateau phase, Zeeman components of the H$\beta$ line are observed, which apparently are formed in the photosphere of the white dwarf. The high difference between the magnetic field of the halo and the accretion spot is noteworthy. \cite{Ferrario15} provides a table of magnetic field definitions contains five polars for which the magnetic field was measured both from cyclotron spectra and from Zeeman absorptions of the halo. For three objects (V834~Cen, MR~Ser, V884~Her), both methods give the same estimates of the magnetic field strength, for one object, both estimates differ slightly (EF~Eri, $B_{cyc}/B_{halo} = 21$~MG$/15$~MG). For the BS~Tri polar, close estimates of the magnetic field in the accretion spot and the halo were also obtained \citep{Kolbin22}. In terms of the behavior of the magnetic field, {\obj} is similar to the BL~Hyi polar, where the magnetic field in the accretion spot is almost twice as strong as the halo magnetic field ($B_{cyc} = 23$~MG and $B_{halo} = 12$~MG; \cite{Ferrario96, Schwope95}).

The dynamic spectrum of the H$\alpha$ line can be described by two components: the main component containing most of the flux ($K\approx 260 $~km/s) and the broad base component ($K\approx 1000 $~km/s). An eclipse of the main source of the H$\alpha$ emission is observed. The flux curve and the radial velocity  curve of the eclipsed H$\alpha$ component agree with the assumption that the line is formed in the accretion stream near the donor surface. Some deviations of the model radial velocity curve from the observed one are apparently related to the poor separation of the line into components due to the low spectral resolution.

Cyclotron radiation formed in the accretion spot is observed in the spectra of the bright phase. The cyclotron spectra were modeled with a simple model of an accretion spot which is uniform in temperature and density. The absence of harmonics of the cyclotron line in the spectra does not allow one to make unambiguous estimates of the parameters of the radiating medium. The agreement between the observed and model spectra is achieved for temperatures $T_e \gtrsim 15$~keV and magnetic fields $B = 15-40$~MG.

{\bf Acknowledgement.}  The study was supported by a grant of the Russian Science Foundation № 22-72-10064, https://rscf.ru/project/22-72-10064/. The authors are grateful to  TUBITAC, IKI, KFU and Academy of Sciences of RT for partial support in the use of the RTT-150 (Russian-Turkish 1.5m telescope in Antalya).


\begin{thebibliography}{16}


\bibitem[\protect\citeauthoryear{Afanasiev, Moiseev}{2011}]{Afan}
\reference{V.L. Afanasiev, A.V. Moiseev}
{\journal{BaltA}{20}{363}{2011}}

\bibitem[\protect\citeauthoryear{Beuermann, Burwitz, et al.}{2020}]{Beuermann20} 
\reference{K. Beuermann, V. Burwitz, K. Reinsch, et al.}
{\journal{\aap}{634}{91}{2020}}

\bibitem[\protect\citeauthoryear{Campbell, Harrison, et al.}{2008}]{Campbell08} 
\reference{R.K. Campbell, T.E. Harrison, A.D. Schwope, et al.} 
{\journal{\apj}{672}{531}{2008}}

\bibitem[\protect\citeauthoryear{Cropper}{1990}]{Cropper90}
\reference{M. Cropper}
{\journal{SSRV}{54}{195}{1990}}

\bibitem[\protect\citeauthoryear{Chanmugam, Dulk}{1981}]{Chan81}
\reference{G. Chanmugam, G.A. Dulk}
{\journal{\apj}{244}{569}{1981}}

\bibitem[\protect\citeauthoryear{van Dokkum}{2001}]{Dokkum}
\reference{P.G. van Dokkum}
{\journal{\pasp}{113}{1420}{2001}}

\bibitem[\protect\citeauthoryear{Ferrario, Bailey, et al.}{1996}]{Ferrario96}
\reference{L. Ferrario, J. Bailey, D. Wickramasinghe}
{\journal{\mnras}{282}{218}{1996}}

\bibitem[\protect\citeauthoryear{Ferrario, de Martino, at al.}{2015}]{Ferrario15}
\reference{L. Ferrario, D. de Martino, B.T. Gänsicke}
{\journal{\ssr}{191}{111}{2015}}

\bibitem[\protect\citeauthoryear{Flannery}{1975}]{Flannery75}
\reference{B.P. Flannery}
{\journal{\mnras}{170}{325}{1975}}

\bibitem[\protect\citeauthoryear{Hellier}{2001}]{Hellier01}
\reference{С. Hellier}
{Cataclysmic Variable Stars (Springer)}

\bibitem[\protect\citeauthoryear{Horne}{1986}]{Horne86}
\reference{K. Horne}
{\journal{\pasp}{98}{609}{1986}}

\bibitem[\protect\citeauthoryear{King, Cannizzo}{1998}]{King98}
\reference{А.R. King, J.K. Cannizzo}
{\journal{\apj}{499}{348}{1998}}

\bibitem[\protect\citeauthoryear{Knigge, Baraffe, et al.}{2011}]{Knigge11}
\reference{С. Knigge, I. Baraffe, J. Patterson}
{\journal{\apjs}{194}{28}{2011}}

\bibitem[\protect\citeauthoryear{Kolbin, Serebryakova, et al.}{2019}]{Kolbin19}
\reference{А.I. Kolbin, N.A. Serebryakova, M.M. Gabdeev, et al.}
{\journal{\astpbull}{74}{80}{2019}}

\bibitem[\protect\citeauthoryear{Kolbin, Borisov}{2020}]{Kolbin20}
\reference{A.I. Kolbin, N.V. Borisov}
{\journal{\astl}{46}{812}{2020}}

\bibitem[\protect\citeauthoryear{Kolbin, Borisov, et al.}{2022}]{Kolbin22}
\reference{A.I. Kolbin, N.V. Borisov, N.A. Serebriakova, et al.}
{\journal{\mnras}{511}{20}{2022}}

\bibitem[\protect\citeauthoryear{Kotze, Potter, et al.}{2015}]{Kotze15}
\reference{E.J. Kotze, S.B. Potter, V.A. McBride}
{\journal{\aap}{579}{77}{2015}}

\bibitem[\protect\citeauthoryear{Kotze, Potter, et al.}{2016}]{Kotze16}
\reference{E.J. Kotze, S.B. Potter, V.A. McBride}
{\journal{\aap}{595}{47}{2016}}

\bibitem[\protect\citeauthoryear{Marsh, Schwope}{2016}]{Marsh16}
\reference{T.R. Marsh, A.D. Schwope}
{\journal{ASSL}{439}{195}{2016}}

\bibitem[\protect\citeauthoryear{Masci, Laher, et al.}{2018}]{masci18}
\reference{F.~Masci, R.~Laher, B.~Rusholme, et al.}
{\journal{\pasp}{131}{995}{2018}}

\bibitem[\protect\citeauthoryear{Schimeczek, Wunner}{2014}]{Schimeczek14}
\reference{C.~Schimeczek, G.~Wunner}
{\journal{\apjs}{212}{26}{2014}}

\bibitem[\protect\citeauthoryear{Schwope, Beuermann, et al.}{1995}]{Schwope95}
\reference{A.D.~Schwope, K.~Beuermann, S.~Jordan}
{\journal{\aap}{301}{447}{1995}}

\bibitem[\protect\citeauthoryear{Schwope, Mantel, et al.}{1997}]{Schwope97}
\reference{A.D. Schwope, K.H. Mantel, K. Horne}
{\journal{\aap}{319}{894}{1997}}

\bibitem[\protect\citeauthoryear{Schwope, Mengel}{1997}]{schwope97b}
\reference{A.D.~Schwope, S.~Mengel}
{\journal{Astronomische Nachrichten}{318}{25}{1997}}

\bibitem[\protect\citeauthoryear{Schwope, Catal{\'a}n, et al.}{2000}]{Schwope00} 
\reference{A.~D. Schwope, M.~S. Catal{\'a}n, et al.} 
{\journal{\mnras}{313}{533}{2000}}

\bibitem[\protect\citeauthoryear{Schwope, Mackebrandt, et al.}{2015}]{Schwope15}
\reference{A.D. Schwope, F. Mackebrandt, B.D. Thinius, et al.}
{\journal{Astronomische Nachrichten}{336}{115}{2015}}

\bibitem[\protect\citeauthoryear{Silber}{1992}]{Silber92}
\reference{A.D. Silber}
PhD thesis, MIT, (1992).

\bibitem[\protect\citeauthoryear{Szkody, Henden, et al.}{2005}]{Szkody05}
\reference{P. Szkody, A. Henden, et al.}
{\journal{\aj}{129}{2386}{2005}}

\bibitem[\protect\citeauthoryear{Wada, Shimizu, et al.}{1980}]{Wada80}
\reference{Wada T., Shimizu A., et al.}
{\journal{PThPh}{64}{}{1986}}

\bibitem[\protect\citeauthoryear{Wickramasinghe, Tuohy, et al.}{1987}]{wickra87}
\reference{D.T. Wickramasinghe, I.R. Tuohy, N. Visvanathan}
{\journal{\apj}{318}{326}{1987}}

\bibitem[\protect\citeauthoryear{Woelk, Beuermann}{1996}]{Woelk96}
\reference{U. Woelk, K. Beuermann}
{\journal{\aap}{306}{232}{1996}}

\bibitem[\protect\citeauthoryear{Warner}{1995}]{Warner95}
\reference{B. Warner}
{Cataclysmic Variable Stars (Cambridge Univ. Press, Cambridge)}

\bibitem[\protect\citeauthoryear{Zorotovic, Schreiber, et al.}{2011}]{Zolot11}
\reference{M. Zorotovic, M.R. Schreiber, B.T. Gänsicke, et al.}
{\journal{\aap}{536}{42}{2011}}









\end{thebibliography}
\end{document}